\documentstyle[prbbib,aps]{revtex}
\tightenlines
\title{Spin-Orbit-Induced Kondo Size Effect
in Thin Films with 5/2-spin Impurities}
\author{O. \'Ujs\'aghy$^a$ and A. Zawadowski$^{a,b}$} 
\address{$^a$Institute of Physics and Research Group of Hungarian Academy
of Sciences, Technical University of Budapest,
H-1521 Budapest, Hungary}
\address{$^b$Research Institute for Solid State Physics, POB 49, H-1525
Budapest, Hungary}
\date{\today}
\begin{document}

\draft
\maketitle

\begin{abstract}
Recently, for spin $S=5/2$ impurities quite different size dependence 
of the Kondo contribution to the resistivity was found experimentally than for
$S=2$. Therefore previous calculation about the effect of the 
spin-orbit-induced magnetic anisotropy on the Kondo amplitude of the 
resistivity is extended to the case of $S=5/2$ impurity spin which differs 
from the integer spin case as the ground state is degenerated. In this case 
the Kondo contribution remains finite      
when the sample size goes to zero and the thickness dependence in 
the Kondo resistivity is much weaker for Cu(Mn). The behavior of 
the Kondo coefficient as a function of the thickness depends on the Kondo 
temperature, that is somewhat stronger for larger 
$T_K$. Comparing our results with a recent experiment in thin Cu(Mn) films,
we find a good agreement. 
\end{abstract}


\section{Introduction}
\label{sec:1}

The Kondo effect \cite{Kondo} in samples with reduced dimensions
(thin films, narrow wires) is one of the most challenging problems
in the field. Most of the experiments 
\cite{kiserletek1} have shown that the Kondo contribution to the 
resistivity is suppressed when the sample size is reduced or the 
disorder in the sample is increased. In addition the different
thermopowers of samples with different thickness gave further
evidences for the existence of size dependence \cite{Strunk}.
The previously examined
samples were Au(Fe), Cu(Fe), and Cu(Cr) alloys, i.e., alloys with 
integer spin impurities. Surprisingly, however, very weak
size dependence has been found recently in Cu(Mn) alloys \cite{Giordano}.
The first possible explanation related to 
the size of the Kondo screening cloud \cite{Bergmann} was ruled out
both theoretically \cite{Affleck} and experimentally \cite{kiserletek2}.
In the limit of strong disorder the experiments might be well explained
with the theory of Phillips and co-workers based on weak localization
\cite{Phillips}. 

In the dilute limit the theory of the spin-orbit-induced magnetic 
anisotropy proposed by the authors \cite{Mi1} was able to explain
every experiment in samples with reduced dimensions, small disorder 
and integer spin impurities for thin layers \cite{Mi2,Borda}. 
Recently an elegant method was developed by Fomin and co-workers
\cite{Fomin} which can be applied for a general geometry.
According to this theory 
\cite{Mi1,Fomin} the 
spin-orbit interaction of the conduction electrons on the non-magnetic host 
atoms can result in a magnetic anisotropy for the magnetic impurity. 
This anisotropy can be described by the Hamiltonian 
$H_a=K_d ({\bf n}{\bf S})^2$ where ${\bf n}$ is the normal direction 
of the experienced surface element, ${\bf S}$ is the spin of the impurity, 
and $K_a$ is the anisotropy constant which is always positive and 
inversely proportional to the distance of the impurity from the surface.   
Due to this anisotropy the spin multiplet splits according to 
the value of $S_z$. In the case of integer spin (e.g., $S=2$ for Fe), the 
lowest level is a singlet with $S_z=0$ thus at a given temperature the 
impurities close enough to the surface, where the splitting is greater 
than $k_B T$, cannot contribute to the Kondo resistivity. 
When the sample size becomes smaller, more and more impurity spins freeze 
out, reducing the amplitude of the Kondo resistivity \cite{Mi2}.

The theory predicts, however, different behavior for samples with 
half-integer impurity spin (e.g., $S=5/2$ for Mn) 
which is recently in the center of interest. In the one-half 
case when the anisotropy looses its meaning
\cite{Mi2}, in fact no anomalous size dependence has been found for
Ce impurities by Roth and co-workers \cite{Roth}. In half-integer $S>1/2$
case the lowest 
level is a doublet ($S_z=\pm 1/2$) thus even an impurity close to the 
surface (large anisotropy) has a contribution to the Kondo resistivity. 
Even accepting that the surface anisotropy reduces the free spin of the 
manganese to a doublet, it is far from trivial that in this case the size
dependence is drastically suppressed, therefore an elaborate theory is 
required.

In this paper we calculate the Kondo resistivity in thin films
of magnetic alloys with $S=5/2$ impurity with helps of the simple model
and MRG calculation of Ref.~\cite{Mi2}. Fitting to the Kondo resistivity
the $\Delta\rho=-B\ln T$ function, we calculate the coefficient
$B$ in terms of the film thickness which is quite different from 
the $S=2$ case. These results will be compared to the recent experiment
by Jacobs and Giordano
\cite{Giordano} presented in thin Cu(Mn) films and we compare 
also the case of alloys with different Kondo temperatures.

\section{Kondo size effect in thin films with impurities S=5/2}
\label{sec:2}

In the presence of the spin-orbit-induced surface anisotropy the 
Kondo Hamiltonian is 
\begin{eqnarray}
H&=&\sum\limits_{k,\sigma}\!\!\varepsilon_k \,a_{k\sigma}^\dagger
a_{k\sigma}+H_a\nonumber \\
&+& \sum\limits_{\scriptstyle k,k^\prime,\sigma,\sigma^\prime \atop
\scriptstyle M,M^\prime}
J_{MM'} 
\bbox{S}_{MM'}\,(a_{k\sigma}^\dagger 
\bbox {\sigma}_{\sigma\sigma'}
a_{k'\sigma'}),
\label{H}\end{eqnarray}
where $a_{k\sigma}^\dagger$ ($a_{k\sigma}$) creates (annihilates)
a conduction electron with momentum $k$, spin $\sigma$ and energy
$\varepsilon_k$ measured from the Fermi level. The conduction electron band
is taken with constant energy density $\rho_0$ for one spin direction,
with a sharp and symmetric bandwidth cutoff $D$.
$\bbox {\sigma}$ stands for the Pauli matrices,
$J_{MM'}$'s are the effective Kondo couplings and 
$H_a=K M^2$ is the anisotropy Hamiltonian when the quantization axis is
parallel to ${\bf n}$.
Applying the Callan-Symantzik multiplicative renormalization group
(MRG) method to the problem, the next to leading logarithmic scaling 
equations for the dimensionless couplings $j_{MM'}=\rho_0 J_{MM'}$ 
were calculated for any impurity spin in Ref.~\cite{Mi2}. 
There was a multiple step scaling performed, corresponding to the freezing 
out of different intermediate states due to the surface anisotropy. 
After exploiting the $j_{M,M'}=j_{M',M}=j_{-M,-M'}$ symmetries of the 
problem, the scaling equations (see Eq.~(20) and (21) in 
Ref.~\cite{Mi2}) were solved numerically 
in terms of the scaling parameter $x=\ln({D_0\over D})$.

The results for $S=5/2$, $j_0=0.0294$, and $D_0=10^5$K, i.e., 
$T_K=10^{-3}$K for Cu(Mn), can be seen in Fig.\ \ref{fig1}. 
We can see from the plot that, when $K/T$ is large enough, at $D=T$ every
coupling can be neglected, except the $j_{{1\over 2}{1\over 2}}$ and
$j_{-{1\over 2}{1\over 2}}$. This corresponds to the freezing out of the
given higher $S_z$ states, but it can be seen well that the two lowest 
energy states are still significant even for large anisotropy.

The Kondo resistivity calculated from the running couplings at $D=T$
by solving the Boltzmann equations (see Ref.~\cite{Mi2}) is
\begin{equation}
\rho_{\text{Kondo}}(K,T)={3\over 4}{m\over e^2}
{2\pi\over\varepsilon_F\rho_0^2}{c\over\int d\varepsilon 
\biggl(-{\partial f_0\over\partial\varepsilon}\biggr) F^{-1}(\varepsilon)}
\label{cond1}\end{equation}
where $c$ is the impurity concentration, $\varepsilon_F$ is the Fermi 
energy, $f_0$ is the electron distribution function in the absence of an
electric field, and $F$ is a function of the running couplings at $D=T$,
spin factors, the strength of the anisotropy $K$, and the temperature, 
defined in Ref.~\cite{Mi2}.
For thin films the Kondo resistivity calculated in the frame of a simple 
model where the two surfaces contribute to the anisotropy constant in an 
additive way as
\begin{equation}                                            
K(d,t)=K_d+K_{t-d}={\alpha\over d}+{\alpha\over t-d},
\label{kate}\end{equation}
is
\begin{equation}
{\bar\rho_{\text{Kondo}}}(t,T)\approx{1\over
t}\int\limits_0^t\rho_{\text{Kondo}}(K(x,t),T) dx
\label{ugyanaz}\end{equation}
where $\alpha$ is the proportionality factor of the spin-orbit-induced
surface anisotropy, $t$ is the thickness of the film and we have used the 
fact that the Kondo contribution to the resistivity is smaller by a factor 
of $10^{-3}$ than the residual normal impurity contribution (see for the 
details Ref.~\cite{Mi2}).

In Fig.\ \ref{fig2} the resistivity as a function of $\ln T$ can be seen 
for different $t/\alpha$ for $S=5/2$, $j_0=0.0294$, and $D_0=10^5$K, i.e.,
$T_K=10^{-3}$K as for Cu(Mn). Thus reducing the thickness of the film for
given $\alpha$, the Kondo contribution to the resistivity is reduced 
comparing to the bulk value, but 
for given thickness there is a temperature below that the reduction turns 
into an increase. Fitting the logarithmic function $-B\ln T$ to the Kondo
resistivity, the plots of the coefficient $B/B_{\text{bulk}}$ as a function 
of the thickness can be seen in Fig.\ \ref{fig3}. Here we can better 
see this very different behavior from the $S=2$ case (cf. Ref.~\cite{Mi2}).
First of all, the Kondo amplitude is reduced comparing to the bulk value, 
but the dependence on the thickness is much weaker than for $S=2$.
Secondly, for small $t/\alpha$'s the coefficient does not go to
zero as for $S=2$, but has a minima at $t/\alpha\sim 6{1\over K}$ and
then changes sign, begins to increase. 
This corresponds to that for an 
$S=5/2$ impurity in the presence of the anisotropy, the lowest energy states
are the $S_z=\pm{1\over 2}$ doublet which give a contribution to the 
resistivity even for large anisotropy (small distance from the surface)
and which can also be larger than the bulk value as a consequence of the
spin factors in the scaling equations. 
Because of the large domain
of the microscopic estimation of the anisotropy constant 
($\alpha\sim 100\text{\AA}K-10^4\text{\AA}K$), we cannot give
a precise microscopic prediction for the place of the minima.  
According to the the fits on the experimental results on Au(Fe) 
\cite{Mi2,Borda} and Cu(Fe) \cite{Mi3}, $\alpha\sim 40-250\text{\AA}K$
from which we can obtain a rough estimation for the place of the
minima as $t_{\it min}\sim 240-1500\text{\AA}$. However, the theoretical
calculation is not reliable in that region where $j_{{1\over 2}{1\over 2}}$
is already in the strong coupling limit. Thus the minima may be a sign
of the breakdown of the weak coupling calculation.

We have fitted in the $T=1.4-3.9$ K temperature regime, thus  
below $4$K where the electron-phonon  
interaction can still be neglected \cite{Giordano} and well above the 
Kondo temperature where the weak coupling approximation is justified, thus 
our perturbative calculation and the logarithmic 
function for the Kondo resistivity is valid. 
These results are in good agreement with the recent experiments of
Jacobs and Giordano \cite{Giordano} on thin films of Cu(Mn).

In Fig.\ \ref{fig4} we have examined what like is the function $B/c$ for
different Kondo temperatures corresponding to different $5/2$-spin alloys. 
It can be seen from the figure that a minima in the 
${B\over c}(t/\alpha)$ function
for small $t/\alpha$'s is present at pretty much the same place (i.e., 
$t/\alpha\sim 5-8{1\over K}$). The character of the $B$ coefficient
comparing to $B_{\rm bulk}$ would roughly the same, but
the absolute measure of the thickness dependence and the reduction from
the bulk value become larger for larger $T_K$ when we fit in the same 
temperature regime corresponding to the larger bulk value (see the
figure caption of Fig.\ \ref{fig4}).

\section{Conclusions}
\label{sec:6}

In this paper we have reexamined our previous calculation \cite{Mi2} about 
the Kondo resistivity in thin films of alloys with $S=5/2$ impurities in 
the presence of spin-orbit-induced surface anisotropy.
First we presented our results on Kondo resistivity for $S=5/2$ and 
$T_K=10^{-3}$K, i.e., for Cu(Mn), and fitted the $-B\ln T$ function on it.
The $B$ function in terms of the film thickness is plotted in  
Fig.\ \ref{fig3} from which we can see that there is a reduction in 
resistivity comparing to the bulk value, but the $B$ coefficient depends 
much weaker on the thickness than in case of $S=2$, and for small 
$t/\alpha$'s $B$ does not go to zero, but it has a minima at ca. 
$t/\alpha=6{1\over K}$ below that it increases which may be only a sign
of the breakdown of the weak coupling calculation. 
These results are in good agreement with 
the recent experiment on Cu(Mn) films \cite{Giordano}. We do not get, 
however, their factor of 30 difference from the bulk value.

Then we have examined what like is the $B/c$ coefficient for different
Kondo temperatures, i.e., for different $5/2$-spin alloys.
We have find that the character of the $B/c$ function is the same,
but the reduction from the bulk value and the thickness dependence is 
somewhat stronger for larger $T_K$. 

Summarizing, the level structure of the impurity is very different
for integer and half-integer spin and at the surface the spin has
a degenerate ground state in the latter case, but we obtained an 
essentially suppressed size dependence which is far from being
trivial. The actually dependence is an interplay between different
effects as the strength of the anisotropy and the different amplitudes 
and temperature
dependence of coupling strength of the electron induced transitions 
between levels. The drastically different behavior for the integer
and half-integer cases found experimentally and determined theoretically
provides a further support of the surface anisotropy as the origin of 
the size dependence.

We would like to thank N. Giordano for helpful discussions.
This work was supported by Grant OTKA $T024005$ and $T029813$. 
One of us (A.Z.) benefited from 
the hospitality of the Meissner Institute and the LMU in Munich and
supported by the Humboldt Foundation.

\begin{figure}
\caption{The running couplings for $S=5/2$ as a function of
$x=\ln{D_0\over D}$ at $K=30$ K, $T=0.3$ K with parameters $D_0=10^5$ K
and $j_0=0.0294$ ($T_K=10^{-3}$ K).}
\label{fig1}
\end{figure}

\begin{figure}
\caption{The resistivity for $S=5/2$ for different values
of $t/\alpha$.
(1) $t/\alpha=3{1\over K}$
(2) $t/\alpha=6{1\over K}$
(3) $t/\alpha=\infty (K=0)$
(4) $t/\alpha=13{1\over K}$
(5) $t/\alpha=40{1\over K}$
The initial parameters were chosen
as $j_0=0.0294$ and $D_0=10^5$ K, $T_K=10^{-3}$ K, and
$\rho^{(0)}={6\pi m c\over 4 e^2\varepsilon_F\rho_0^2}$.}
\label{fig2}
\end{figure}

\begin{figure}
\caption{The coefficient $B/B_{\rm bulk}$ as a function
of $t/\alpha$ for $S=5/2$ and $T_K=10^{-3}$ K.}
\label{fig3}
\end{figure}

\begin{figure}
\caption{The calculated coefficient $B_{\rm calc}={B\over c}
{6\pi m\over 4 e^2\varepsilon_F\rho_0^2}$ as a function of $t/\alpha$
for $S=5/2$ for different Kondo temperatures.
(1) $T_K=10^{-1}$ K, $B_{\rm calc}^{{\rm bulk}}=0.1793$;
(2) $T_K=10^{-2}$ K, $B_{\rm calc}^{{\rm bulk}}=0.0377$;
(3) $T_K=10^{-3}$ K, $B_{\rm calc}^{{\rm bulk}}=0.0127$.}
\label{fig4}
\end{figure}

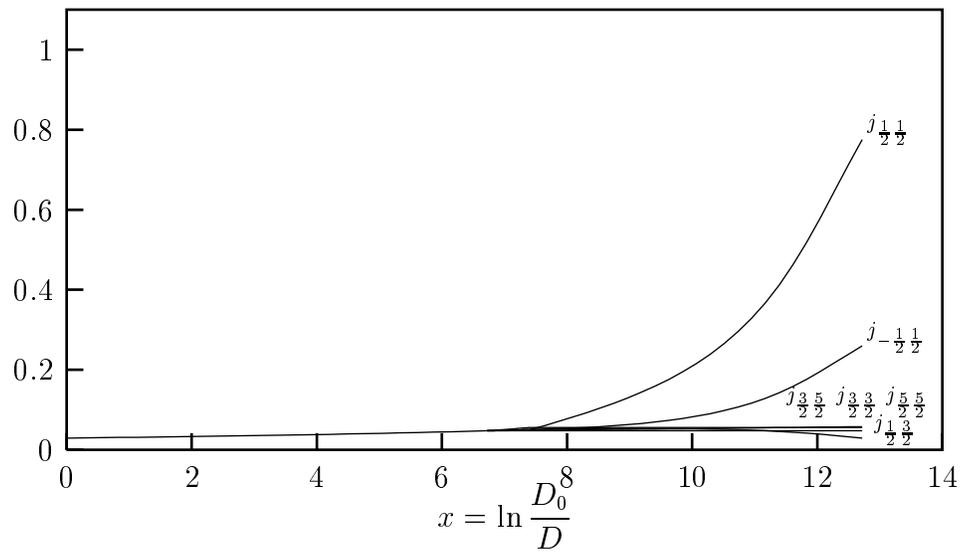
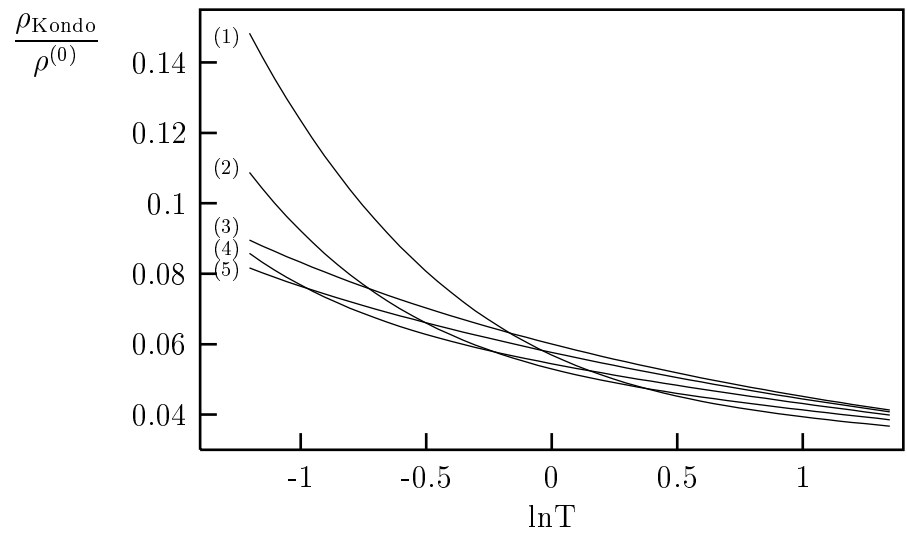
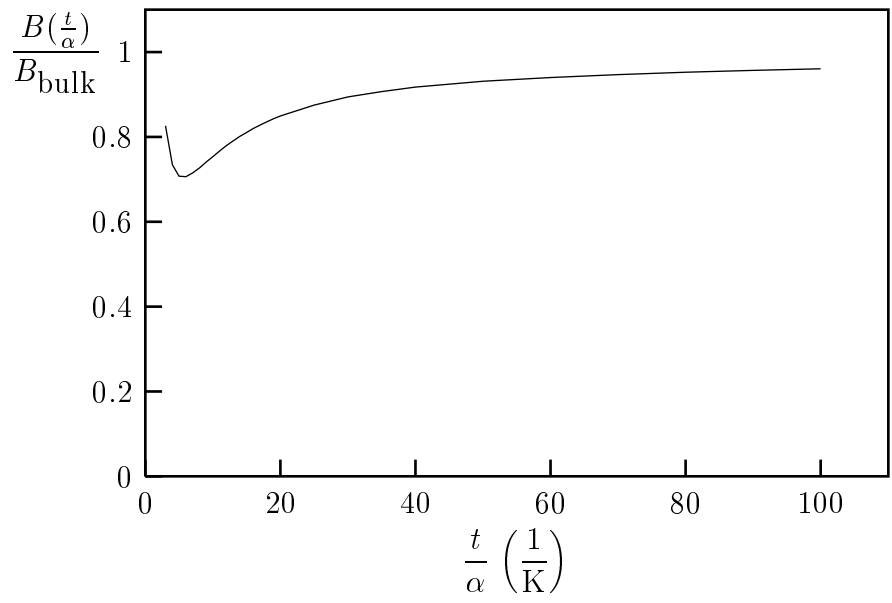
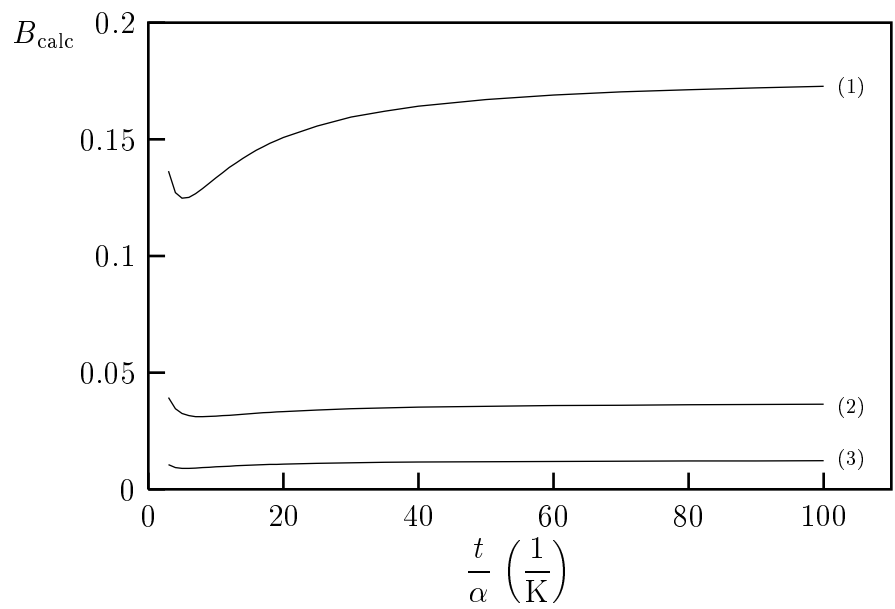
\end{document}